# Title:

Gradual emergence of temporal structures depending on the distance between neighboring callers in natural habitat of male treefrogs

## Authors

Ikkyu Aihara[1,*], Ryu Takeda[2], Masahiro Shirasaka[3], Daichi Kominami[4], Hiromitsu Awano[5,6], and Masayuki Murata[4]

## Affiliation:

1. Division of Information Engineering, Institute of Systems and Information Engineering, University of Tsukuba, Ibaraki, 305-8573, Japan
2. SANKEN, The University of Osaka, Ibaraki, Osaka, 567-0047, Japan
3. Department of Computer Science, Graduate School of Systems and Information Engineering, University of Tsukuba, Ibaraki, 305-8573, Japan
4. Graduate School of Information Science and Technology, The University of Osaka, Osaka 565-0871, Japan
5. Graduate School of Informatics, Kyoto University, Kyoto, 606-8501, Japan
6. Graduate School of Engineering, Nagoya University, Nagoya, 464-8601, Japan

## Corresponding author:

Ikkyu Aihara (aihara@cs.tsukuba.ac.jp)

## Abstract

Collective behavior, such as in-phase and anti-phase synchronizations in acoustic animals, is one of the most striking examples of self-organization in animal communication. However, understanding how synchronization emerges in natural environments has been limited by the difficulty of tracking multiple callers simultaneously in the wild. Here, we quantified spatio-temporal frequency characteristics in the choruses of wild treefrogs by combining large-scale sound-imaging and microphone-array systems. First, we performed field recordings by combining the sound-imaging system (25-66 units of sound-imaging devices) and microphone-array system (16-24 channels of microphones) between 2021 and 2023. Second, we analyzed the video and audio data and quantified the call frequency, call timing, and caller position of each male across 11 choruses comprising 66 males in total. Based on this dataset, we have shown that synchronized behavior gradually emerges between neighboring callers depending on their distances even when the call frequency and chorus density moderately vary. These findings provide field-based evidence for how local acoustic interactions generate collective temporal order in animal groups.

# Introduction

Males of acoustic animals (e.g., insects and frogs) show well-organized temporal structures in their choruses through interactions with other individuals. For instance, male frogs usually alternate their calls with neighbors (Aihara et al., 2011; Aihara et al., 2014; Brush and Narins, 1989; Gerhardt and Huber, 2002; Grafe, 1996; Jones et al., 2014; Wells, 2007). This alternating behavior is described as call alternation characterized by a phase difference of 180° (Aihara, 2009; Aihara et al., 2011, 2014; Greenfield et al., 2021, 2023, 2026). Playback experiments have indicated that call alternation allows males to avoid acoustic interference and effectively advertise themselves to conspecific females (Schwartz, 1987; Legett et al., 2019). In contrast, male insects (e.g., katydids, crickets and cicadas) show either call alternation (a phase difference of 180°) or call synchrony (a phase difference of 0°) depending on the species (Gerhardt and Huber, 2002; Sismondo, 1990; Greenfield and Roizen, 1993; Greenfield et al., 2021, 2023, 2026; Cooley and Marshall, 2001; Ishimaru and Aihara, 2025). Playback experiments have indicated that call synchrony allows males to mask the acoustic signals of competitors and reduce their attractiveness to females (Cooley and Marshall, 2001; Greenfield and Roizen, 1993). Thus, well-organized temporal structures are common in the choruses of acoustic animals and play an important role in mating with females.

In this study, we test the hypothesis that the degree of temporal organization is associated with the spatial arrangement of callers and the difference in call frequencies in the wild. First, the acoustic signals of males play important roles in the advertisement of territories as well as mate attraction (Gerhardt and Huber, 2002; Wells, 2007), generating complicated spatial distributions in the wild. Given that sounds attenuate as the distance between callers increases, the inter-frog distance can affect the precision of synchronized behavior, such as call alternation and call synchrony. In addition, previous studies have indicated that well-organized temporal patterns among neighboring callers disappear as chorus size increases (Schwartz, 1993; Schwartz et al., 2002). Thus, the occurrence of temporal structures can also be affected by chorus size although the need to avoid call overlaps may increase in a larger chorus (or a denser chorus). Second, acoustic signals with similar frequencies can be difficult to distinguish from one another particularly when they overlap in time. In many frog species, call frequency depends on body size and thus represents an individual-specific characteristic (Arak 1988; Wollerman 1998; Giacoma et al., 1997; Gwynne and Bailey 1988). Accordingly, calls with more similar frequencies may be more susceptible to acoustic interference, potentially increasing the need to avoid temporal overlap. In natural choruses, the degree of acoustic interference between neighboring callers is expected to depend on both naturally occurring differences in call frequencies and their spatial arrangement, which in turn could influence their temporal relationships. However, previous field studies on the temporal chorus structures of frogs and insects had relatively small sample sizes, limiting statistical analyses that account for variation in call frequencies as well as spatial arrangement. Recent advances in the technologies of audio recordings and signal processing allow us to estimate detailed features of acoustic signals and the positions of callers (Aihara et al., 2011, 2014; Au and Herzing, 2003; Fujioka et al., 2016; Grafe, 1997; Jones et al., 2014; Rhinehart et al., 2020; Simmons et al., 2008; Suzuki et al., 2018; Calsbeek et al., 2022). The application or extension of these technologies to the choruses of insects and frogs should contribute to further understanding of the origin of temporal chorus structures in the wild.

Japanese treefrogs (*Hyla japonica*; *Dryophytes japonicus*; *Dryophytes leopardus*) distribute widely in Japan (Matsui and Maeda, 2018). Male Japanese treefrogs aggregate and produce calling bouts, which consist of a series of advertisement calls, in unison (Aihara et al., 2019). The calls are periodically produced at a relatively shorter interval of about 0.3 sec in each bout (Aihara, 2009; Aihara et al., 2011, 2014). Despite such a dense and acoustically complicated environment, male Japanese treefrogs exhibit well-organized temporal structures in their choruses. Laboratory experiments demonstrated that two males synchronized in anti-phase (Aihara, 2009) while three males showed tri-phase synchronization and clustered anti-phase synchronization (Aihara et al., 2011). Field recordings also documented anti-phase synchronization between neighbors in the wild (Mizumoto et al., 2011; Aihara et al., 2014; Aihara et al., 2016). However, (1) the lack of datasets with sufficient sample sizes capturing variations in spatial distributions and chorus densities and (2) the lack of information on frequency features especially in field sites precluded us from identifying the origin of temporal structures in the context of the spatio-frequency features in natural choruses. In this study, we have carried out field recordings on 11 nights between 2021 and 2023 and used the resulting dataset to test our hypothesis that the degree of temporal organization is associated with the spatial arrangement of callers and the difference in their call frequencies in the wild.

## Materials and Methods

Field observations have demonstrated that, at the beginning of the breeding season, male Japanese treefrogs aggregate along the edges of rice paddies (Mizumoto et al., 2011; Aihara et al., 2014; Aihara et al., 2016). To monitor such a wide area containing multiple sound sources, we have combined two recording systems: (1) the sound-imaging system and (2) the microphone-array system. The sound-imaging system consists of a device that is illuminated when detecting nearby sounds (Mizumoto et al., 2011; Awano et al., 2021). In previous studies, we have deployed dozens of the devices along the edges of rice paddies and estimated the position of callers and the times of their calls (Mizumoto et al., 2011; Aihara et al., 2014; Awano et al., 2021). Although the sound-imaging system allows us to quantify the spatio-temporal structures of frog choruses, it does not allow us to estimate the frequencies of acoustic signals.

Figure 1 schematically illustrates our recording system that was set up in a natural environment. First, we deployed 25-66 units of the sound-imaging device along the edge of a rice paddy to cover the one-dimensional distributions of male treefrogs and recorded the illumination patterns of the devices using a video camera (SONY, HDR-PJ630) at a frame rate of 60 frames per second (fps). The response frequency of the sound-imaging devices was tuned to the dominant frequency of the calls of the focal species (*H. japonica*; *D. japonicus*; *D. leopardus*) in advance, following the procedure proposed in a previous study (Awano et al., 2021). Second, we deployed two or three units of the microphone array (System in Frontier Inc., TAMAGO; each microphone array consists of eight microphones) to estimate the frequencies of acoustic signals produced by male frogs. We connected each microphone array to a laptop computer and recorded audio signals at a sampling frequency of 16 kHz. There was vegetation around the rice paddy that could disturb the propagation of audio signals between the sound sources and microphones. To minimize this negative effect, we deployed the microphone arrays inside the rice paddy at almost the same height as the calling frogs by using miniature tripods (Ulanzi, MT-08). These recordings were carried out on 11 nights between 2021 and 2023 (June 2, 10, 11, 14, 16, and 18, 2021; June 17 and 24, 2022; and June 14, 16, and 30, 2023) at a paddy field in Ishioka City, Ibaraki, Japan (36° 18'01.5"N 140° 10'32.7"E).

We quantified the positions of callers as well as the times and frequencies of their calls from the video and audio data. First, we analyzed the video data following the methods proposed in previous studies (Mizumoto et al., 2011; Aihara et al., 2014; Awano et al., 2021) and estimated the positions of the devices [pixel] and the times [sec] when the LED of each device was activated by the calls of male treefrogs. The coordinates of each device [pixel] were converted to those in an actual space [cm] according to the method called *homography* (Solem, 2012; Aihara et al., 2017) by using four reference points that had been measured by a laser distance meter (MAX Co. Ltd., LS-811) in advance. Consequently, we quantified the positions of callers and the times of their calls for a total of 66 males that stayed along the edge of the paddy field and consistently called during the field recordings. Second, we applied the method of sound-source separation to the audio data of the microphone arrays. We selected the sections containing the signals from multiple callers based on the illumination patterns of the sound-imaging devices, extracted the corresponding audio data of the microphone arrays, and carried out the sound-source separation using independent vector analysis (Kim et al., 2006), a well-established signal processing method for blind source separation (see Supplementary Information for details). Finally, we calculated the power spectra of the separated audio signals and estimated the dominant frequencies of calls for all the identified callers (66 males across 11 nights in total). In this study, we assume that the dominant frequences are individual-specific and do not change through interactions with other individuals.

To quantify the strength of synchronization for each pair of callers, we have calculated a phase difference from the sequences of call times. The phase difference between the $n^{\text{th}}$ and $m^{\text{th}}$ frogs is defined by the following equation (Pikovsky et al., 2001; Aihara et al., 2011, 2014; Greenfield et al., 2021):

$$\phi_{nm}(j) = 2\pi \frac{t_m(k) - t_n(j)}{t_n(j+1) - t_n(j)}, \qquad (1)$$

where

$$t_n(j) \le t_m(k) < t_n(j+1). \qquad (2)$$

Here, $t_n(j)$ represents the time of the $j^{\text{th}}$ call produced by the $n^{\text{th}}$ frog and $t_m(k)$ represents the time of the $k^{\text{th}}$ call produced by the $m^{\text{th}}$ frog. The condition in Equation (2) restricts the range of $\phi_{nm}(j)$ between 0 and $2\pi$. Given the strong periodicity during calling behavior (Aihara et al., 2011 and 2014) and the intermittent switching between calling behavior and a silent state (Aihara et al., 2019), we have calculated this quantity when the inter-call interval satisfies $t_n(j+1) - t_n(j) < 0.5$ sec, a threshold slightly longer than a typical inter-call interval in this species (Aihara, 2009; Aihara et al., 2011, 2014). This phase difference allows us to discriminate the types of temporal structures depending on its value. For example, in-phase synchronization (i.e., call synchrony) is described by $\phi_{nm}(j) = 0$ [rad] (or 0°) while anti-phase synchronization (i.e., call alternation) is described by $\phi_{nm}(j) = \pi\ [rad]\ (or\ 180°)$. Because Japanese treefrogs tend to synchronize in anti-phase (Aihara, 2009; Aihara et al., 2011, 2014 and 2016), the distribution of the phase difference is expected to be localized around $\pi$.

To quantify the sharpness of the distribution of the phase difference, we define the following quantity according to the methodology of circular statistics (Fisher, 1993; Kanji, 1993):

$$\gamma_{nm} \exp(i\psi_{nm}) = \frac{1}{N} \sum_j \exp(i\phi_{nm}(j)). \qquad (3)$$

Here, $\gamma_{nm}$ and $\psi_{nm}$ are the length and angle of the mean vector for the sequence of the phase difference $\phi_{nm}(j)$; $N$ is the number of data samples. For example, $\gamma_{nm}$ takes the maximum value of 1 when the focal pair of calling males show perfect synchronization and all the data of the phase difference $\phi_{nm}(j)$ are the same; $\gamma_{nm}$ is close to zero when the focal pair do not show any type of synchronization and the data of $\phi_{nm}(j)$ are randomly distributed. To obtain a reliable mean vector from Equation (3), we restricted our analysis to the pairs of males for which more than 300 samples of the phase difference were obtained.

Finally, we have statistically examined the origin of temporal structures in frog choruses. In particular, we have used the following statistical model (the generalized linear mixed model (Schall 1991; Faraway 2006)):

$$\gamma_{nm} \sim Beta(\mu_{nm}\delta, (1-\mu_{nm})\delta), \qquad (4)$$

and

$$logit(\mu_{nm}) = a_{distance}X_{distance,nm} + a_{freq}X_{freq,nm} + a_{density}X_{density,nm} + a_0 + \xi_{date} + \xi_n + \xi_m. \qquad (5)$$

Here, $\gamma_{nm} \in [0,1]$ is the length of the mean vector of the phase difference that quantifies the strength of synchronization (Equation (3)). Since the beta distribution ranges from 0 to 1 like $\gamma_{nm}$, we assume that $\gamma_{nm}$ follows the beta distribution with the mean $\mu_{nm}$ and the precision parameter $\delta$ (Equation (4)). The mean $\mu_{nm}$ is related to the explanatory variables through the logit link function (Equati on (5)). This framework is consistent with beta regression (Ferrari et al., 2004), a well-established regression method. The explanatory variables $X_{distance,nm}$, $X_{freq,nm}$ and $X_{density,nm}$ are the distance [m], the difference in call frequencies [kHz] and the chorus density [Num. of callers/m] for the pair of the $n^{th}$ and $m^{th}$ frogs, respectively. The chorus density $X_{density,nm}$ has been calculated as $N_{caller}/L_{dist}$, where $L_{dist}$ is the total length of the one-dimensional distribution of the sound-imaging devices and $N_{caller}$ is the number of calling males along that line. In Equation (5), we introduce two kinds of random effects associated with the date ($\xi_{date}$) and the identities of the two males included in the focal pair ($\xi_n$ and $\xi_m$), respectively. The random effect $\xi_{date}$ is assumed to follow a Gaussian distribution with mean 0 and standard deviation $\sigma_{date}$ while $\xi_n$ and $\xi_m$ follow a Gaussian distribution with mean 0 and standard deviation $\sigma_{frog}$. The explanatory variables $X_{distance,nm}$, $X_{freq,nm}$ and $X_{density,nm}$ were normalized between 0 and 1, allowing us to compare the relative magnitudes of the coefficients $a_{distance}$, $a_{freq}$ and $a_{density}$. The unknown parameters of this model (i.e., $a_{distance}$, $a_{freq}$, $a_{density}$, $a_0$, $\delta$, $\sigma_{date}$ and $\sigma_{frog}$) were estimated by Bayesian inference based on our empirical data of $\gamma_{nm}$, $X_{distance,nm}$, $X_{freq,nm}$ and $X_{density,nm}$ obtained from a total of 66 male treefrogs across 11 nights.

## Results

The spatio-temporal frequency characteristics of frog choruses have been quantified by our methodology. Figure 2A and B shows the spatial distribution and illumination patterns of the sound-imaging devices, respectively. The devices activated by the calls are highlighted by pink markers (Figure 2A), corresponding to the positions of calling males. The illumination patterns of the devices capture the periodic calling behavior of individual males (Figure 2B), which is consistent with previous reports on this species (Matsui and Maeda, 2018; Aihara, 2009; Aihara et al., 2011, 2014, 2016). Figure 2C shows a representative result of the sound source separation; the comparison of the separated audio signals with the illumination patterns of the sound-imaging devices allows us to identify which device has responded to each audio signal. Finally, we have estimated the dominant frequency of the calls for each male by using the separated audio signals.

Figure 3 shows examples of spatio-temporal frequency characteristics in choruses of male treefrogs. These are examples of small choruses with three males staying along the edge of a rice paddy. In the first example (Figure 3A), neighboring males were located at similar distances of about 2 m. The dominant frequencies of their calls varied moderately among 3.06, 3.33 and 3.39 kHz. The phase differences between neighbors were localized around $\pi$, demonstrating the occurrences of anti-phase synchronization. The second example (Figure 3B) indicates the change in temporal chorus patterns depending on datasets. In particular, the spatial distribution was more asymmetric in this case; the distance between the 1st and 2nd frogs was much shorter than that between the 2nd and 3rd frogs. Accordingly, the distribution of the phase difference between the closer pair (i.e., the 1st and 2nd frogs) was more sharply peaked than that between the distant pair (i.e., the 2nd and 3rd frogs), indicating the effect of spatial distributions on temporal chorus structures.

Figure 4 shows the time series data of phase differences obtained from the dataset of Figure 3A. In Figure 4A, the phase differences between neighboring callers fluctuate around $\pi$, corresponding to anti-phase synchronization within both neighboring pairs. By contrast, in Figure 4B, the phase difference between Frogs 1 and 2 fluctuates around $\pi$ whereas that between Frogs 2 and 3 fluctuates around 0, corresponding to a different type of clustered anti-phase synchronization (i.e., anti-phase synchronization between Frog 1 and the other frogs that exhibit in-phase synchronization). Although the chorus spontaneously switches between these states, anti-phase synchronization between neighboring callers (Figure 4A) is dominant in our datasets. This feature explains why the distributions of phase differences are centered around $\pi$ (see the bottom panels of Figure 3).

The comprehensive analyses of our datasets obtained from a total of 66 male frogs on 11 nights allow us to quantify variations in both spatial arrangements and dominant frequencies. The distances between neighbors distributed in a one-dimensional pattern along the edge of the paddy field varied considerably (mean $\pm$ std $= 2.97 \pm 2.04$ m, N=56 pairs), whereas the chorus size (the number of male frogs participating in the choruses during each recording) ranged from three to nine (mean $\pm$ std $= 6.00 \pm 2.10$ callers, N=11 recordings). The distribution of dominant frequencies shows an obvious peak around 3.32 kHz (mean $\pm$ std $= 3.32 \pm 0.17$ kHz, N=66 males; see Figure S3 in Supplementary Information), which is consistent with a previous report on this species (Matsui and Maeda, 2018).

Finally, we have performed statistical analyses using the generalized linear mixed model (Equations (4) and (5)) to examine whether spatial and frequency features are associated with the robustness of temporal chorus patterns. Figure 5 summarizes the results of our analyses for two kinds of datasets: (1) the pairs of neighboring callers ($N = 81$ pairs) and (2) the pairs of non-neighboring callers ($N = 93$ pairs). Here, we classify pairs of callers as neighboring pairs if no other callers are located between them along the edge of the rice paddy, and as non-neighboring pairs if at least one other caller is located between them. As for the first dataset, the 95% confidence interval (CI) of $a_{distance}$ is negative whereas the 95% CIs of $a_{density}$ and $a_{freq}$ include zero (Figure 5(A1)). The scatter plot with the estimated parameters shows an obvious negative relationship between the inter-frog distance and the degree of synchronization (Figure 5(A2)). We have also confirmed that the mean phase difference is close to $\pi$ when $\gamma_{nm}$ is high, demonstrating the occurrences of anti-phase synchronization between close neighbors (see Figure S4 in Supplementary Information). The results of the second dataset (i.e., the pairs of non-neighboring callers) are similar to those of the first dataset. Namely, the 95% CI of $a_{distance}$ is negative whereas the 95% CIs of $a_{density}$ and $a_{freq}$ include zero (Figure 5(B1)). However, the absolute value of $a_{distance}$ is much smaller than that of the first dataset and the upper limit of the 95% CI of $a_{distance}$ is close to zero. This feature indicates that the effect of the distance between callers is significant but much weaker and marginal in the pairs of non-neighboring callers than in the pairs of neighboring callers. The results of the parameter estimation for all the unknown parameters (i.e., $a_{distance}$, $a_{freq}$, $a_{density}$, $a_0$, $\Phi$, $\sigma_{date}$ and $\sigma_{frog}$) are shown in Figure S5 of Supplementary Information.

## Discussion

In this study, we examined spatio-temporal frequency characteristics in the choruses of wild treefrogs (*H. japonica*; *D. japonicus*; *D. leopardus*). First, we performed field recordings by combining the sound-imaging system and the microphone-array system. Second, we analyzed the video and audio data and quantified the positions of callers as well as the times and frequencies of their calls. Our analyses using the statistical model showed that (1) the distance between neighboring callers was significantly associated with the strength of synchronization even when accounting for the difference in call frequencies and chorus density and (2) a similar result was obtained for the pairs of non-neighboring callers but the effect of distance was much weaker and marginal.

Here, we discuss the biological implications of this study. Previous studies examined the spatio-temporal structures of choruses in other frog species through field recordings. These studies demonstrated that anti-phase synchronization (call alternation) was common in the choruses of male frogs (Brush and Narins, 1989; Grafe, 1996; Jones et al., 2014; Aihara et al., 2014), but statistical analyses based on large datasets accounting for variations in chorus size and call frequency were not performed. In addition, previous studies indicated that the well-organized structure between neighbors disappeared when the chorus size became larger (Schwartz, 1993; Schwartz et al., 2002). This study demonstrated that synchronized behavior (anti-phase synchronization) emerged between close neighbors even when the chorus density varied moderately, indicating a high ability for acoustic communication among wild treefrogs in a natural environment. In particular, our finding that anti-phase synchronization becomes robust as the distance between neighboring callers decreases (Figure 5 (A2)) may reflect selective attention toward nearest neighbors (Brush and Narins, 1989; Jones et al., 2014; Greenfield and Rand, 2000; Greenfield and Snedden, 2003) and reduced attention toward more distant callers. On the other hand, a significant relationship was not detected between the difference in call frequencies and the strength of synchronization even in the pairs of neighboring callers (Figure 5(A1))). Given that dominant frequencies of acoustic signals are generally similar in the same species (Gerhardt and Huber, 2002), we speculate that the relatively small variation in dominant call frequencies among male Japanese treefrogs (see Figure S3) results in strong interference in the frequency domain and thereby induces alternating calling behavior that allows males to temporally avoid call overlaps.

In this study, we have quantified the spatio-temporal frequency characteristics in the choruses of wild treefrogs by combining two recording systems: the sound-imaging system and the microphone-array system. The point is that our methodology has achieved sound-source separation and localization in natural frog choruses producing intense acoustic interference from similar sound sources distributed across a wide area. Our methodology for the sound-source separation and localization in natural environments may be applicable to various systems such as the choruses of other species of frogs and insects (Gerhardt and Huber, 2002; Wells, 2007) and human conversations (Nagira and Mizumoto, 2023; Hashimoto et al., 2025). Despite this potential applicability, the advantages of our methodology over related methodologies require further evaluation. For instance, an acoustic camera is an alternative approach for sound-source separation and localization in natural environments. An acoustic camera is often equipped with a single video camera and a microphone array, allowing researchers to locate and separate several sound sources. Field recordings with an acoustic camera have demonstrated that the calling behavior of several male frogs can be simultaneously visualized (Calsbeek et al., 2022). However, it has been pointed out that the spatial arrangements of the camera and sound sources (e.g., the distance from sound sources to an acoustic camera and the distance among the sources) substantially affect the quality of sound-source localization and separation (Merino-Martínez et al., 2019; Salom et al., 2023). In particular, greater distances from sound sources to the camera can reduce the signal-to-noise ratio at the input of the microphone array, resulting in the deterioration of the quality of the localization and separation. In contrast, our system consists of spatially distributed sound-imaging devices and microphone arrays, allowing us to choose devices close to sound sources with a high signal-to-noise ratio; this can be advantageous for reliably monitoring wide areas involving spatially distributed sound sources like calling animals.

Here, we discuss the limitations and future extensions of this study. In this study, we have shown the nonlinear relationship between the inter-frog distance and the magnitude of synchronization (Figure 5A). First, the mechanism underlying this relationship warrants further investigation. In particular, our previous model based on a system of coupled mobile oscillators (Aihara et al., 2014) assumes that the coupling strength decreases nonlinearly with the inter-frog distance (i.e., an inverse-square dependence on the inter-frog distance due to sound attenuation), which is consistent with our finding in this study. Given that the mechanism underlying spatio-temporal structures in coupled mobile oscillators has been extensively studied from a theoretical point of view (Tanaka, 2007; O'Keeffe et al., 2017, 2022), future directions of this study include the integration between empirical data and mathematical modeling such as the identification of unknown parameters of a phase oscillator model from the time series data of phases (Ota et al 2020). Second, the effect of spatial arrangements of male frogs on temporal chorus structures remains to be further investigated. In the statistical analyses with the GLMM, we treat the distance between callers as an explanatory variable. However, the relative sound levels of neighboring callers may also affect the strength of their interactions. For instance, if a close neighbor produces much louder calls than a distant neighbor, the louder calls can make it difficult for the focal male to recognize the calls of the distant neighbor. Thus, the threshold for recognizing the calls of other males may depend on the loudness of neighboring callers. This possibility has not yet been examined in the present study. Third, multistability of temporal chorus structures warrants further investigation. As shown in Figures 3 and 4, anti-phase synchronization between neighboring callers appears to be dominant in our datasets. In other words, other synchronization states such as a different type of clustered anti-phase synchronization (Figure 4B) are not dominant. In contrast, our laboratory experiments demonstrated that such states (e.g., the different types of clustered anti-phase synchronization and tri-phase synchronization) occurred more often. The reason for this difference may be that the distance between neighboring males in the wild (approximately 2.97 m) was much larger than that in the laboratory experiments (0.50 m). We speculate that this larger distance in the wild drastically reduced the strength of interactions between non-neighboring callers.

## Ethics statement

This study complied with the relevant guidelines of the University of Tsukuba. Field recordings were conducted on privately owned farmland with the permission of the landowner. No animals were captured, handled, or experimentally manipulated during the study. Because the study involved only non-invasive field recordings and no collection or handling of animals, no additional fieldwork permits or ethical approval were required. s

## Use of generative AI

During the preparation of this manuscript, the authors used ChatGPT (OpenAI) to assist with English phrasing and language refinement. The authors reviewed and edited all AI-generated suggestions and take full responsibility for the content of the manuscript.

# Figure

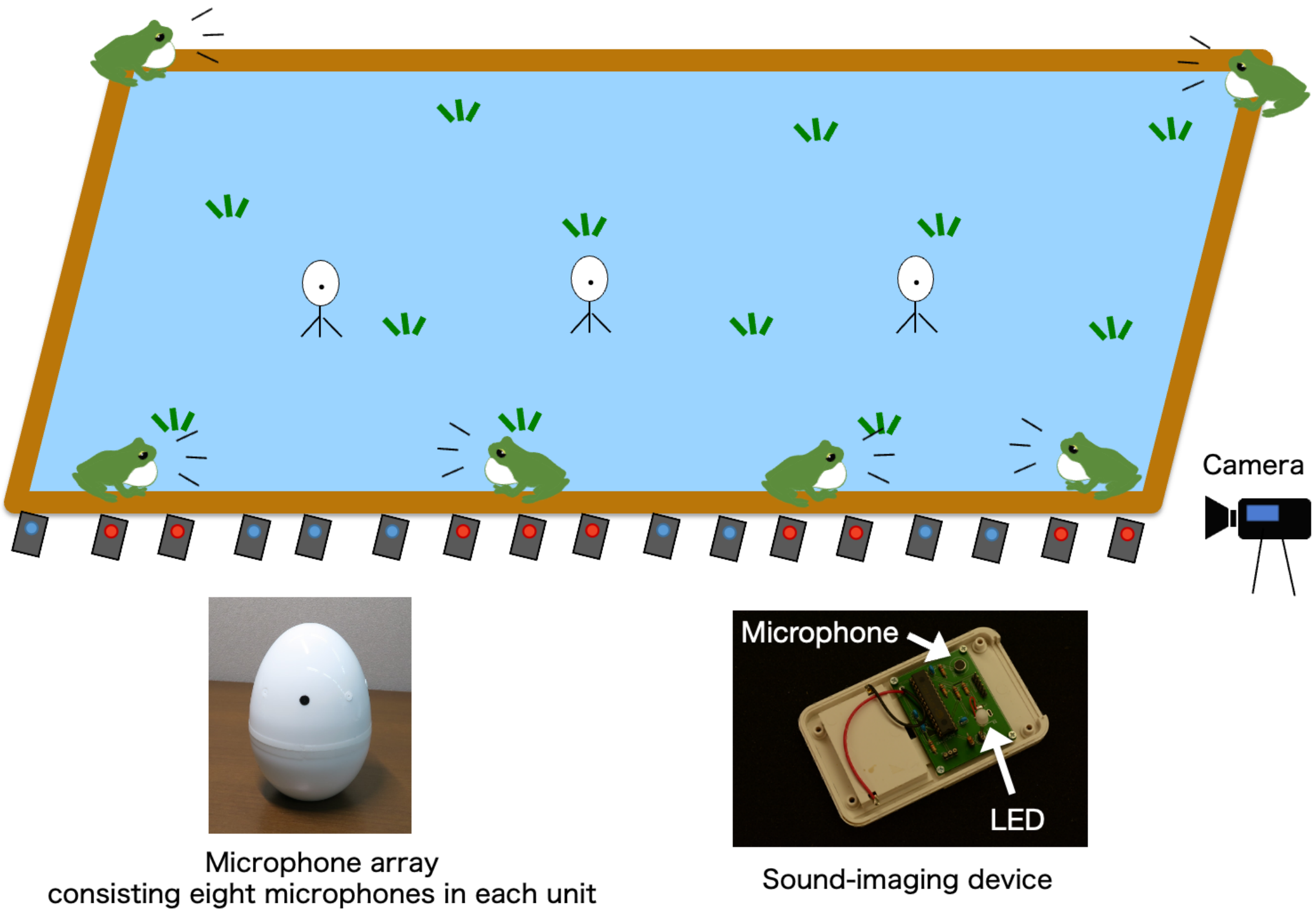

Figure 1: Schematic diagram of our recording system consisting of sound-imaging devices and microphone arrays. Because of the one-dimensional distribution of male Japanese tree frogs (*Hyla japonica*; *Dryophytes japonicus*; *Dryophytes leopardus*), we deployed the sound-imaging devices along the edge of a rice paddy. To record the audio signals of treefrogs with a high signal-to-noise ratio, we spatially distributed multiple microphone arrays inside the rice paddy at almost the same height as the calling frogs. These recordings were carried out on 11 nights between 2021 and 2023, enabling us to detect a total of 66 callers.

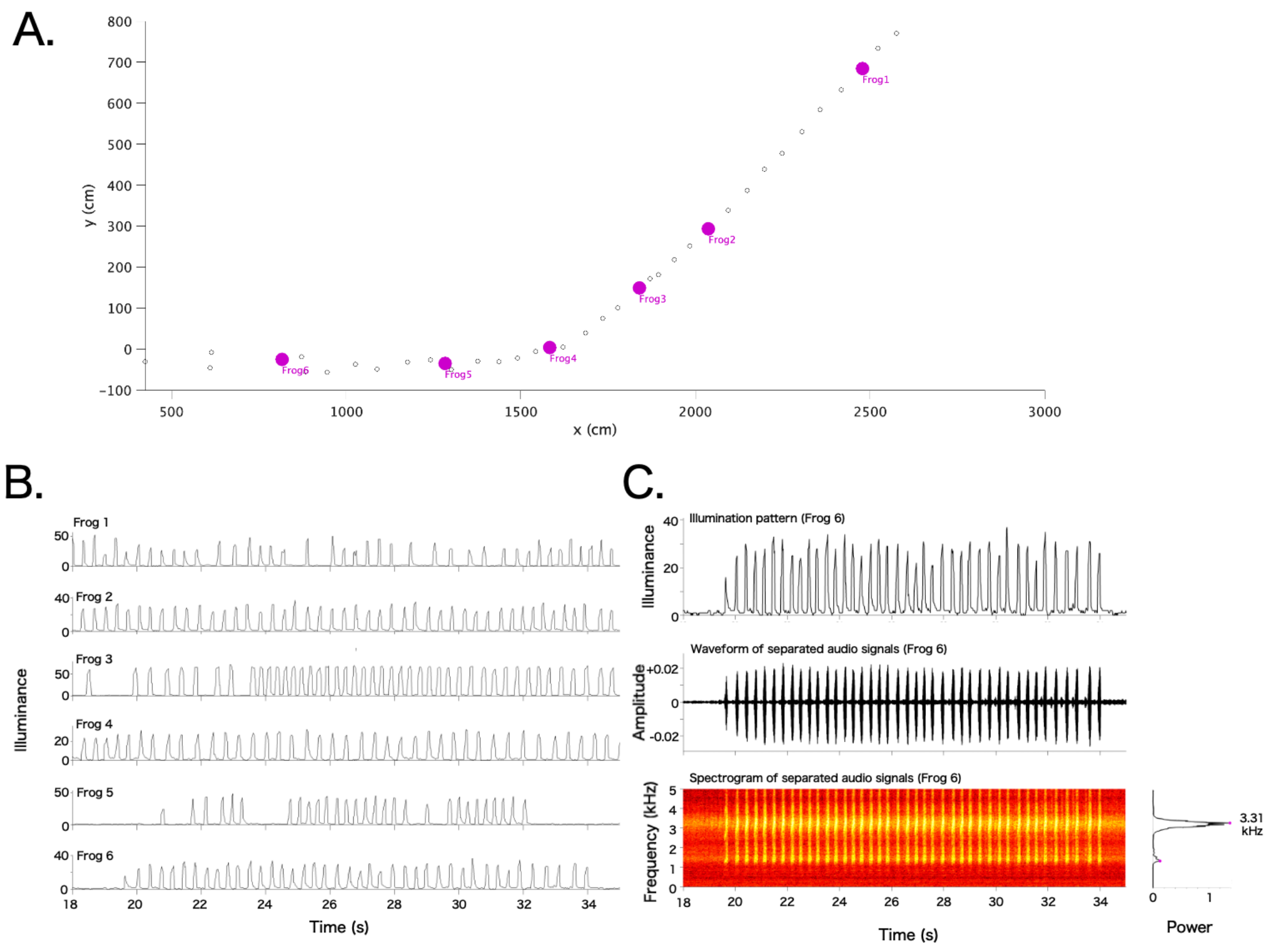

Figure 2: Quantification of spatio-temporal frequency characteristics in choruses of wild treefrogs. (A) The positions of sound-imaging devices estimated from video recordings. The devices activated by the calls of male treefrogs are indicated by pink markers, whereas the other devices are indicated by open black markers. (B) Illumination patterns of the sound-imaging devices that responded to the calls of male treefrogs. The time series data of illuminance allowed us to estimate the time when each call was produced. (C) Sound-source separation from audio recordings. We identified the separated audio signals (middle panel) corresponding to the illumination pattern of the focal device (top panel) by comparing the time evolutions of both signals. Finally, we calculated the power spectrum of the separated audio signals and estimated the dominant call frequency for each male frog (bottom panel).

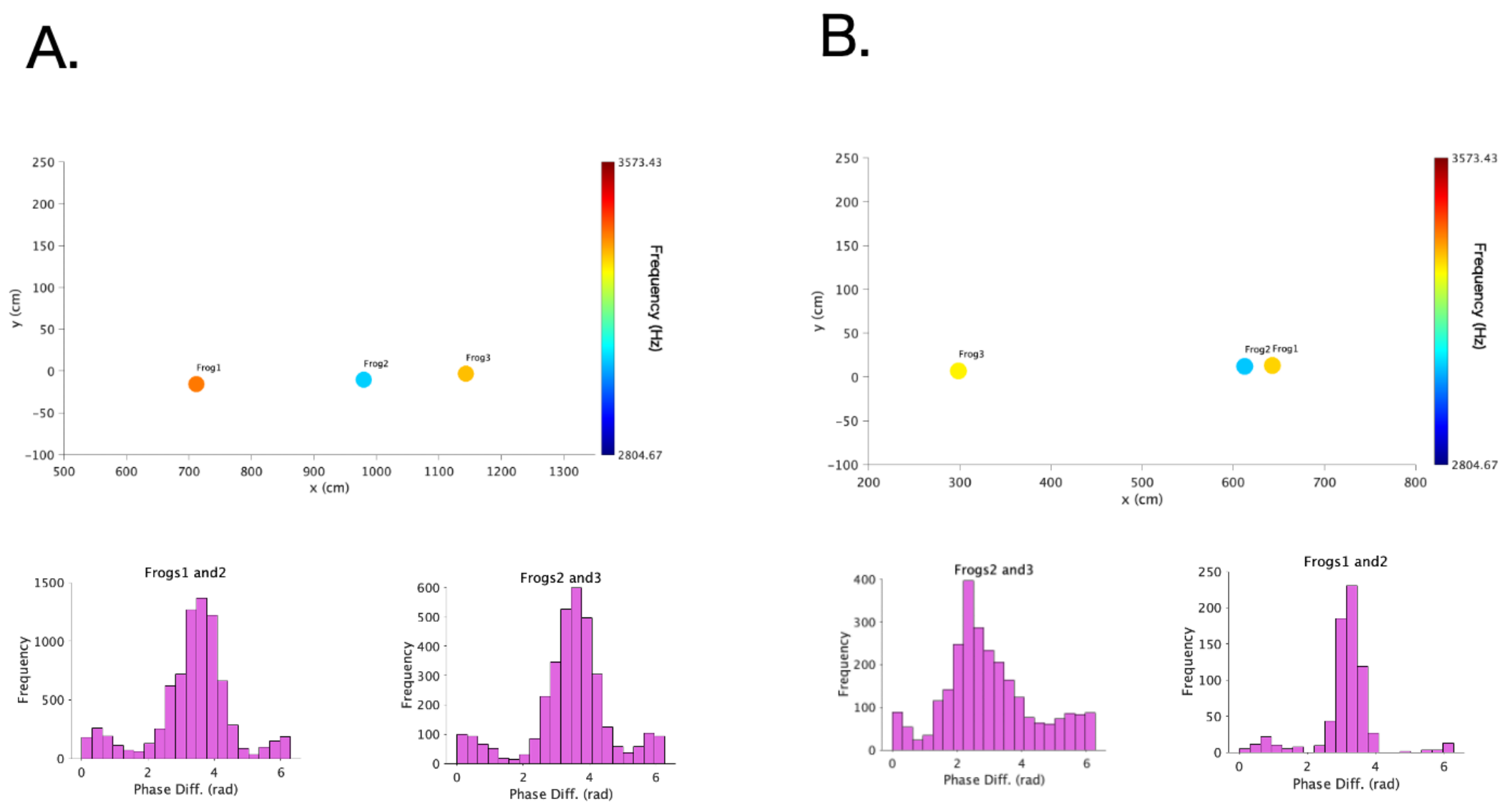

Figure 3: Representative spatio-temporal frequency characteristics in frog choruses. These are two examples of relatively small choruses with three males. Colored markers indicate the positions of callers, together with the dominant frequencies of their calls. The histograms show the distributions of the phase differences between neighbors. In the chorus of Figure 3A, the distance between the 1st and 2nd frogs is close to that between the 2nd and 3rd frogs; the distributions of the phase differences are similar. In contrast, the spatial structure is more asymmetric in the chorus of Figure 3B; the distribution of the phase difference between the closer neighbors (the 1st and 2nd frogs) is more sharply peaked than that between the more distant neighbors (the 2nd and 3rd frogs).

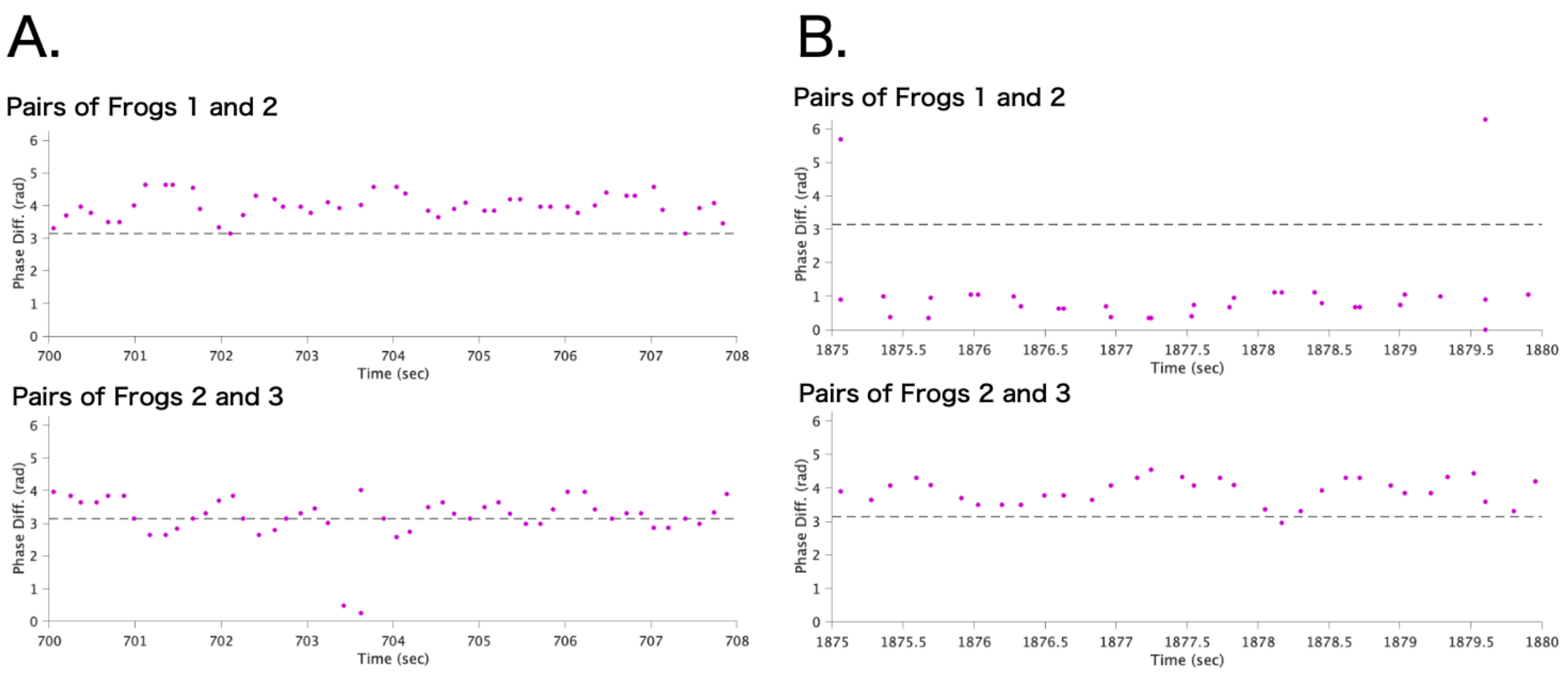

Figure 4: A dynamical feature of phase differences obtained from the dataset of Figure 3A. In Figure 4A, the phase differences between neighboring callers fluctuate around $\pi$, corresponding to anti-phase synchronization between the neighboring pairs. By contrast, in Figure 4B, the phase difference between Frogs 1 and 2 fluctuates around $\pi$ whereas that between Frogs 2 and 3 fluctuates around 0. This phenomenon corresponds to a different type of clustered anti-phase synchronization (i.e., anti-phase synchronization between Frog 1 and the other frogs that exhibit in-phase synchronization).

A1. The pairs of neighboring callers

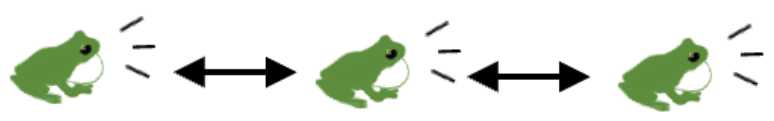

| Parameter | Posterior Mean | 95% CI | Rhat |
|---|---|---|---|
| $a_{distance}$ | -5.62 | -7.54 ~ -3.77 | 1.00 |
| $a_{freq}$ | +0.06 | -0.78 ~ +0.87 | 1.00 |
| $a_{density}$ | -0.59 | -1.59 ~ +0.38 | 1.00 |
| $a_0$ | -0.04 | -0.59 ~ 0.52 | 1.00 |

N = 81 pairs

B1. The pairs of non-neighboring callers

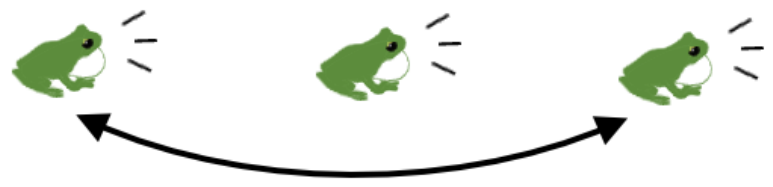

| Parameter | Posterior Mean | 95% CI | Rhat |
|---|---|---|---|
| $a_{distance}$ | -0.90 | -1.71~ -0.11 | 1.00 |
| $a_{freq}$ | -0.13 | -0.74 ~ +0.46 | 1.00 |
| $a_{density}$ | -0.26 | -1.54 ~ +1.00 | 1.00 |
| $a_0$ | -2.06 | -2.64 ~ -1.48 | 1.00 |

N = 93 pairs

A2.

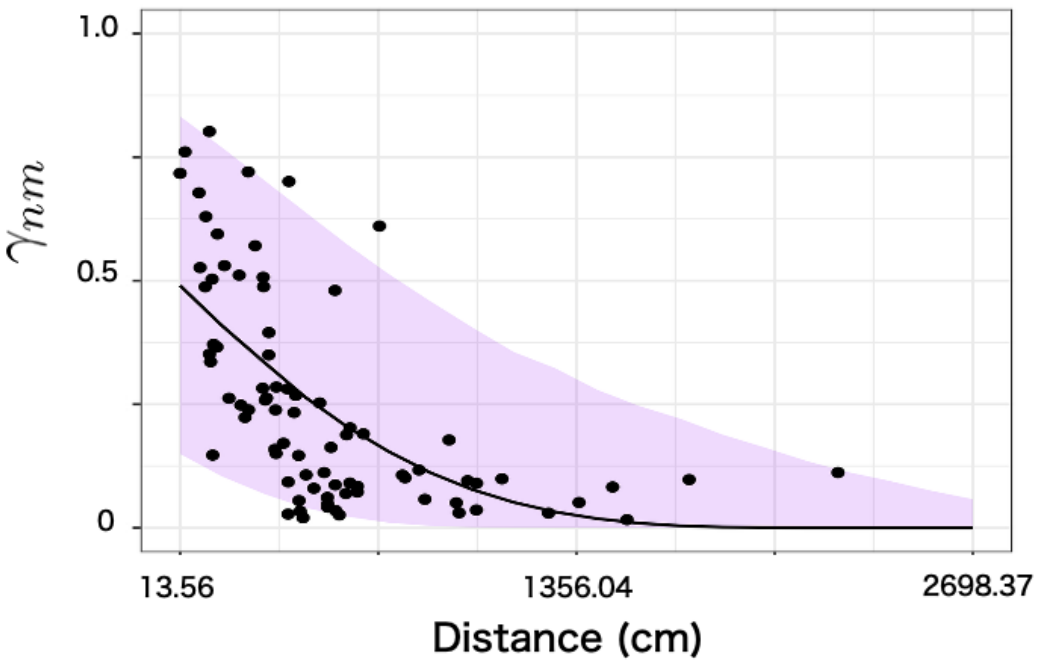

B2.

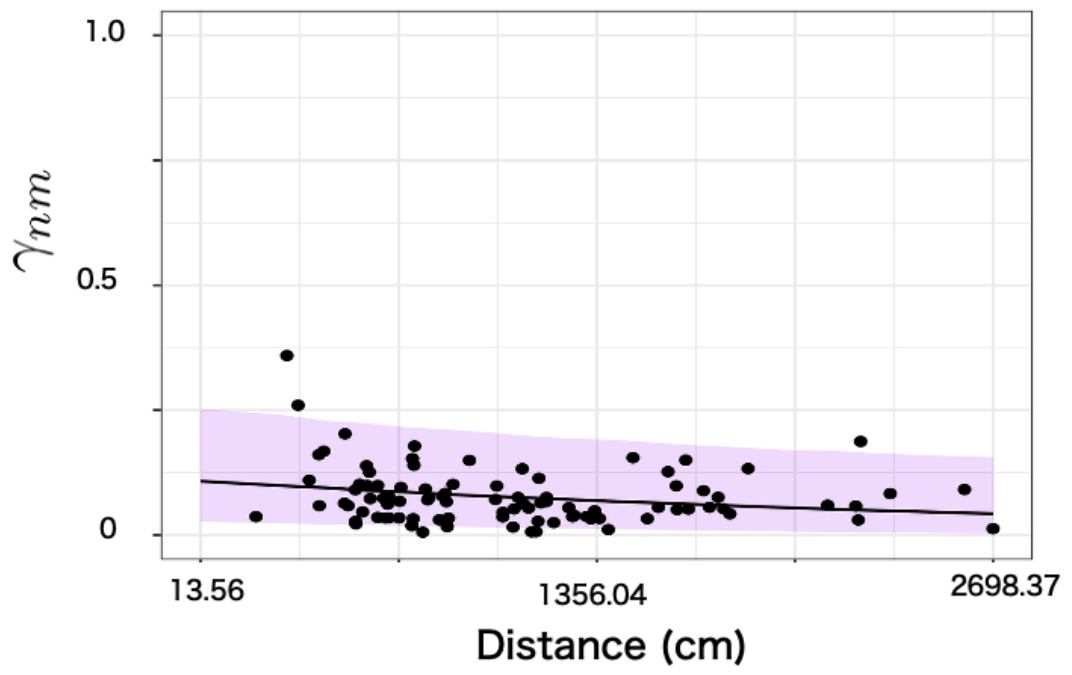

Figure 5: Statistical analyses of the spatio-temporal frequency characteristics using two kinds of datasets: (A) the pairs of neighboring callers ($N = 81$) and (B) the pairs of non-neighboring callers ($N = 93$). Figure 5 (A1) and (B1) shows the parameters of the generalized linear mixed model estimated from respective datasets, whereas Figure 5 (A2) and (B2) shows the scatter plots of empirical data with the estimated models. In Figure 5 (A2) and (B2), the black lines and purple regions represent the regression curves based on the estimated model parameters and their 95% CIs, respectively. The horizontal axes are rescaled to show the actual distance (cm), although the parameter estimation has been carried out using normalized variables (see Materials and Methods for details).

# Supplementary Information

## Sound-source separation

Blind source separation (BSS) is a signal processing method that separates an observed mixture of source signals into isolated ones. We need only the recorded audio data in multi-channel: the positions of microphones and sound sources are not required in advance while traditional beamformers do. The assumption of the method is a statistical property of source signals, such as independence among sound sources, that will satisfy almost all sound sources in the real world. The advantage on the few assumptions makes BSSs a useful tool of sound signals for the non-experts of signal processing.

We simply explain the mixing process between the observed signals at microphones and sound sources. We denote the sound signals captured at each microphone $i$ as $\boldsymbol{x}_t = [x_{t1}, \dots, x_{ti}, \dots, x_{tM}]$ where $t$ represents a discrete time. $\boldsymbol{s}_t = [s_{t1}, \dots, s_{tj}, \dots, s_{tN}]$ represent a sound source vector whose element $s_{tj}$ is a $j$-th source signal at time $t$. The sound wave propagates from sound sources to microphones with power decay and superposition via multiple paths. Such process can be modeled as

$$\boldsymbol{x}_t = \sum_d \boldsymbol{H}_d \boldsymbol{s}_t \tag{S1}$$

where $\boldsymbol{H}_d$ represents a so-called transfer function matrix of acoustic signal. Because we have observed signals from time $t_a$ to $t_b$ (for several seconds), we can utilize statistical information of sound signals.

Our purpose is to estimate the source vector $\boldsymbol{s}_t$ from the observed vector $\boldsymbol{x}_t$. The independent vector analysis (IVA) is a kind of BSSs in the short-time Fourier transformation (STFT) domain (Kim et al., 2006; see Figure S1 for the schematic diagram of IVA). In the STFT domain, the convolution can be approximated by the multiplication of transfer function $\boldsymbol{H}_w$ and source spectrum $\boldsymbol{s}_{wf}$, where $w$ and $f$ mean frequency-bin and frame indices of spectrum, respectively.

$$\boldsymbol{x}_{wf} = \boldsymbol{H}_w \boldsymbol{s}_{wf}, \tag{S2}$$

The estimated source vector $\hat{\boldsymbol{s}}_{wf}$ can be obtained by multiplying separation matrix $\boldsymbol{W}_w$ estimated by IVA to the observed vector $\boldsymbol{x}_{wf}$ as follows:

$$\hat{\boldsymbol{s}}_{wf} = \boldsymbol{W}_w \boldsymbol{x}_{wf}, \qquad \text{(S3)}$$

The time-domain signals corresponding to the separated vectors are resynthesized through inverse STFT.

The estimation of separation filter $\boldsymbol{W}_w$ is achieved by formulating as an optimization problem. Major formulations are the maximization of the likelihood or minimization of KL-divergence. Common factor is the design of the probabilistic density function (PDF) of sound sources. IVA uses a PDF of source vector along the frequency-axis, which overcomes a so-called permutation problem in the STFT processing of independent component analysis (ICA). Please see the detailed estimation algorithm in (Kim et al., 2006).

# Figure legend

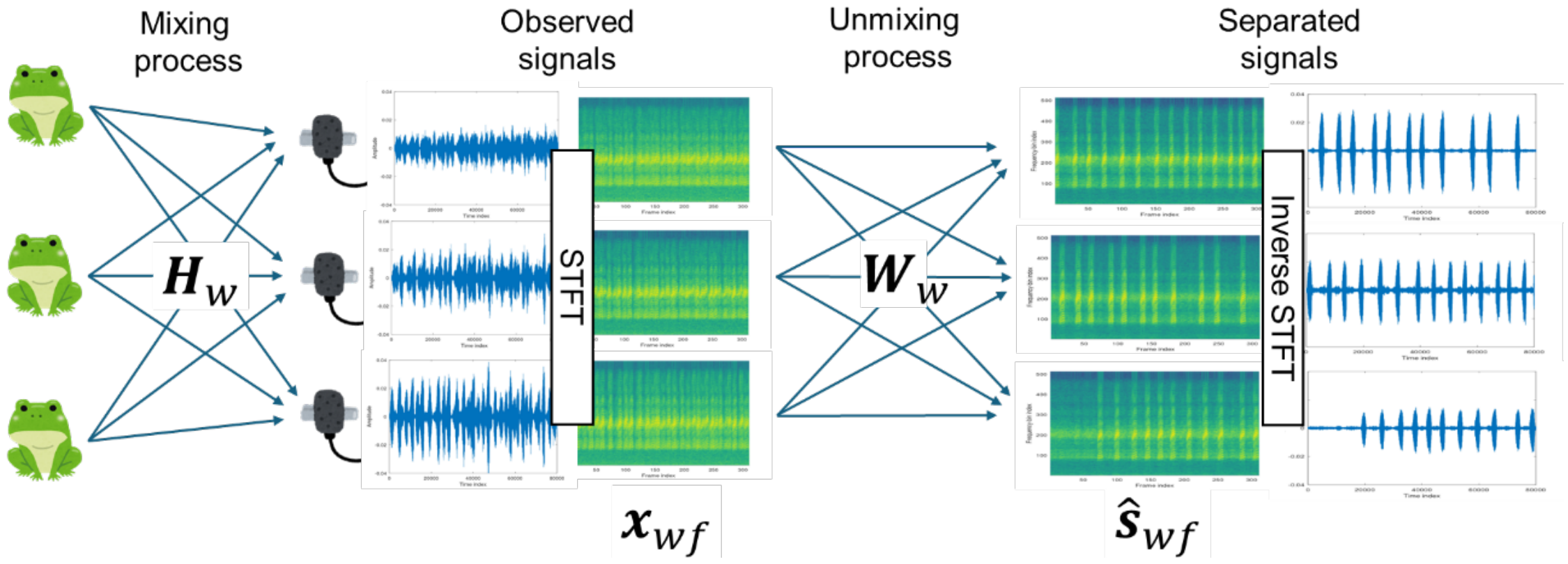

Figure S1: Schematic diagram of the independent vector analysis (IVA) on the assumption of multiple sound sources and microphones.

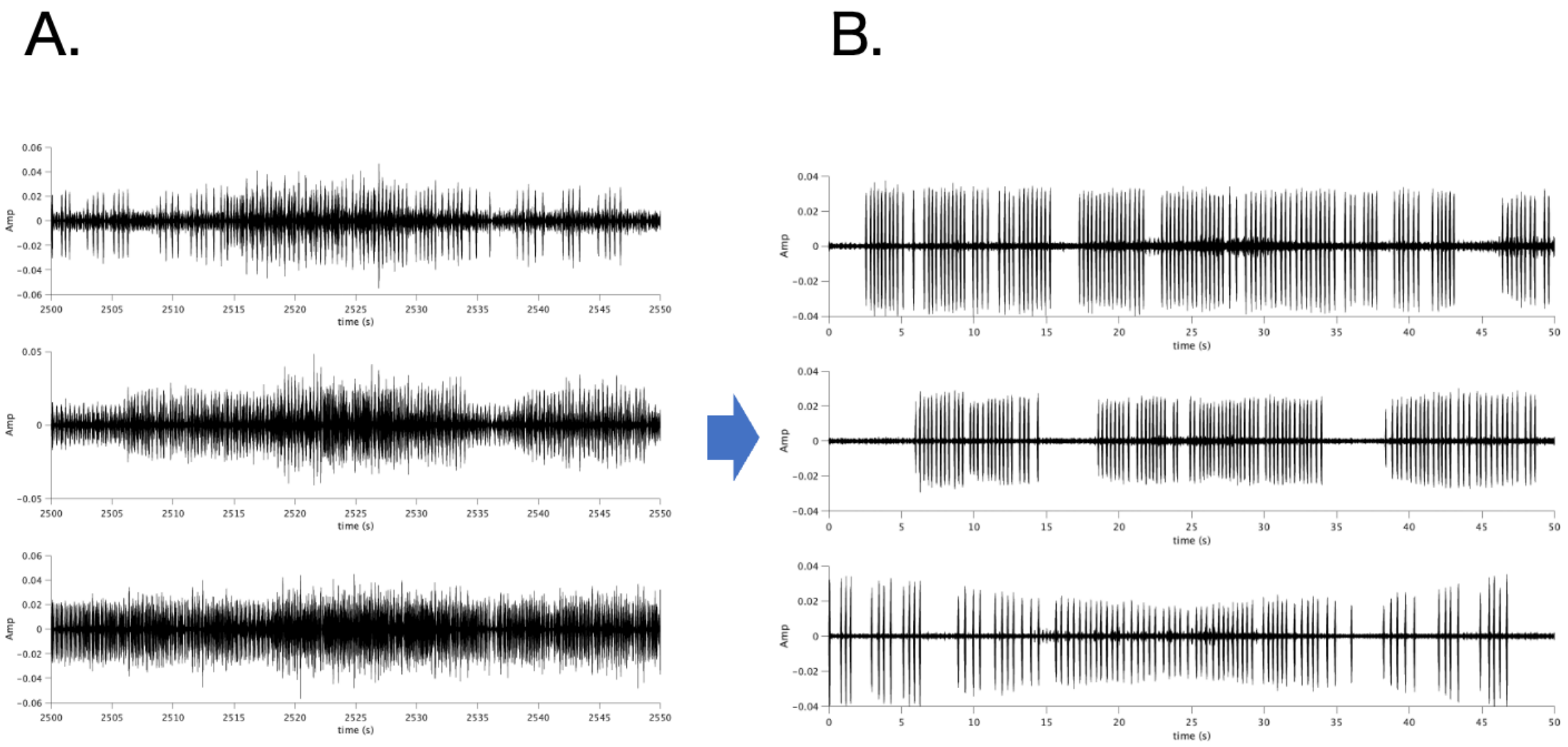

Figure S2: Representative result of sound-source separation. (A) Original audio data obtained from microphone arrays. (B) Separated audio signals obtained using independent vector analysis (IVA). The separated audio signals show strong periodicity with inter-call intervals of approximately 0.3 sec, which is consistent with previously reported calling behavior in this species (Matsui and Maeda, 2018; Aihara, 2009; Aihara et al., 2011, 2014, 2016).

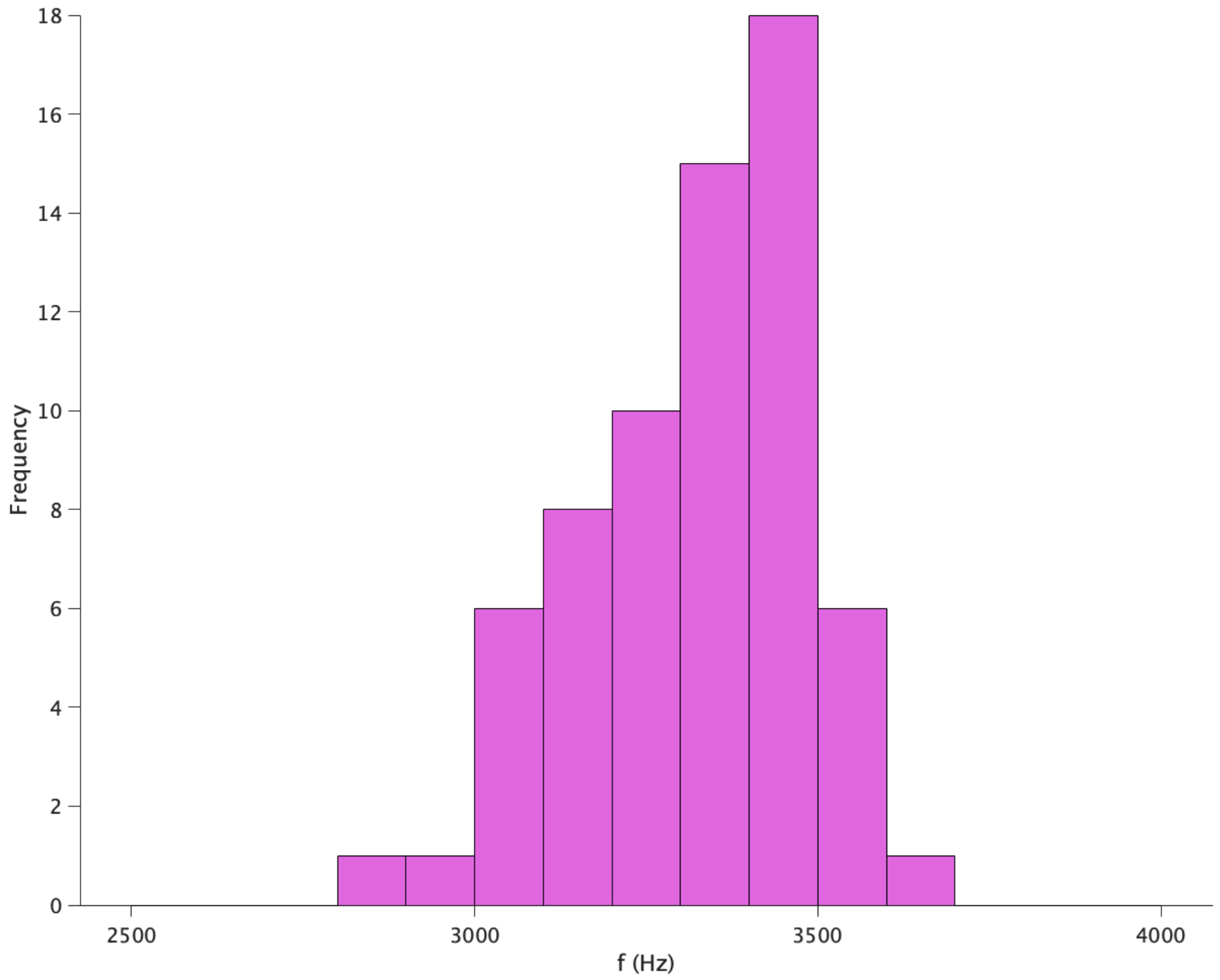

Figure S3: The histogram of dominant call frequencies of male Japanese treefrogs (N=66 males in total). There is an obvious peak around 3.4 kHz, which is consistent with the dominant call frequency previously reported for this species (Matsui and Maeda, 2018).

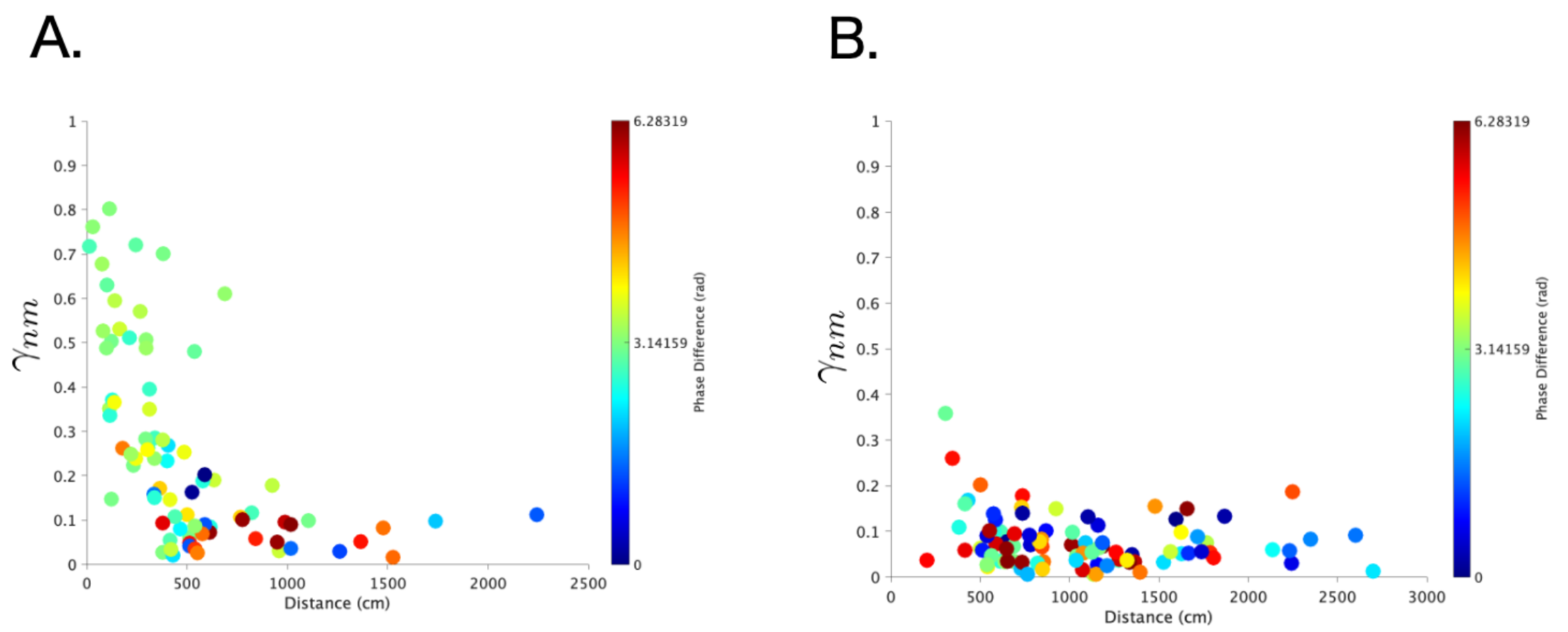

Figure S4: The scatterplots of inter-frog distances and the strengths of synchronization using two kinds of datasets: (A) the pairs of neighboring callers and (B) the pairs of non-neighboring callers. A color gradient represents the values of the mean phase differences. The scatterplots show that the mean phase differences are close to $\pi$ when the strengths of synchronization $\gamma_{nm}$ are high.

**A.** The pairs of neighboring callers

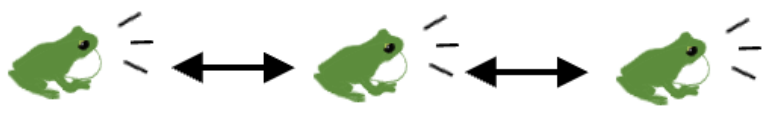

| Parameter | Posterior Mean | 95% CI | Rhat |
|---|---|---|---|
| $a_{distance}$ | -5.62 | -7.54 ~ -3.77 | 1.00 |
| $a_{freq}$ | +0.06 | -0.78 ~ +0.87 | 1.00 |
| $a_{density}$ | -0.59 | -1.59 ~ +0.38 | 1.00 |
| $a_0$ | -0.04 | -0.59 ~ 0.52 | 1.00 |
| $\delta$ | 8.13 | 5.41 ~ 11.93 | 1.00 |
| $\sigma_{date}$ | 0.32 | 0.03 ~ 0.79 | 1.00 |
| $\sigma_{frog}$ | 0.16 | 0.02 ~ 0.42 | 1.00 |

**B.** The pairs of non-neighboring callers

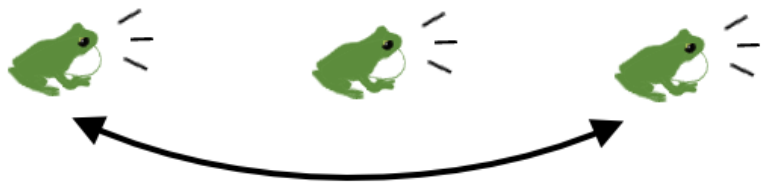

| Parameter | Posterior Mean | 95% CI | Rhat |
|---|---|---|---|
| $a_{distance}$ | -0.90 | -1.71~ -0.11 | 1.00 |
| $a_{freq}$ | -0.13 | -0.74 ~ +0.46 | 1.00 |
| $a_{density}$ | -0.26 | -1.54 ~ +1.00 | 1.00 |
| $a_0$ | -2.06 | -2.64 ~ -1.48 | 1.00 |
| $\delta$ | 39.62 | 26.13 ~ 58.32 | 1.00 |
| $\sigma_{date}$ | 0.50 | 0.12 ~ 1.05 | 1.00 |
| $\sigma_{frog}$ | 0.20 | 0.03 ~ 0.40 | 1.00 |

Figure S5: The results of parameter estimation for all the unknown parameters of the generalized linear mixed model (Equations (4) and (5) in the main manuscript). The parameters $\sigma_{date}$ and $\sigma_{frog}$ represent the standard deviations of Gaussian distributions that the random factors $\xi_{date}$ and $\xi_n$ follow, respectively.